\documentclass[12pt]{article}
\usepackage{myart}

\normalbaselines
\input tcilatex

\begin{document}

\title{Statement of problem on vortical inviscid flow of barotropic and
incompressible fluids.}
\author{Yuri A.Rylov}
\date{Institute for Problems in Mechanics, Russian Academy of Sciences,\\
101-1, Vernadskii Ave., Moscow, 119526, Russia.\\
email: rylov@ipmnet.ru\\
Web site: {$http://rsfq1.physics.sunysb.edu/\symbol{126}rylov/yrylov.htm$}\\
}
\maketitle

\begin{abstract}
The question what information is necessary for determination of a unique
solution of hydrodynamic equations for ideal fluid is investigated.
Arbitrary inviscid flows of the barotropic fluid and of incompressible fluid
are considered. After integrating hydrodynamic equations, all information on
the fluid flow is concentrated in dynamic equations in the form of
indefinite functions, whereas the initial and boundary conditions contain
information on the fluid particle labeling. It is shown that for
determination of the stationary flow of the incompressible fluid the
vorticity on any stream line must be given. Giving the velocity on the
boundary, one does not determine the vorticity, in general. If there are
closed stream lines, the vorticity cannot be given on them via boundary
conditions. This circumstance explains existence of different stationary
vortical flows under the same boundary conditions.
\end{abstract}


\section{Introduction}

In the present paper we consider the statement of the flow problem of
barotropic fluid. If compressibility of the fluid tends to zero, we obtain
an incompressible fluid. We consider the incompressible fluid as a special
case of the barotropic fluid, when its compressibility tends to zero. In the
passage to limit the dynamic equations, describing evolution of density $%
\rho $ and that of the velocity potential $\varphi $, lose temporal
derivatives and turn into constraints on the state of the incompressible
fluid. 
\begin{equation}
\partial _{0}\rho +\mathbf{\nabla }\left( \rho \mathbf{v}\right)
=0\;\;\;\rightarrow \;\;\;\mathbf{\nabla v}=0  \label{h1.0}
\end{equation}
As a result the statement of the flow problem appears to be different for
barotropic fluid and for the incompressible one. On one hand, the
description of the incompressible fluid is simpler, than that of barotropic
one. On the other hand, the incompressible fluid is a nonphysical fluid,
because the speed of sound is infinite, and constraints on the state of the
incompressible fluid appear to be nonphysical constraints. As a result the
statement of the flow problem for the incompressible fluid appears to be
complicated, than for the barotropic one.

Nonstationary flows are too difficult for calculations, and as a rule one
considers stationary flows, which do not contain temporal derivatives. This
fact complicates statement of the flow problem, because the problem cannot
be considered to be an evolutional problem. Finally, the rotational
stationary flows are too difficult for calculation also, and one considers
usually stationary irrotational flows of the incompressible fluid. Statement
of the problem for rotational flows of the incompressible fluid and for the
irrotational ones appear to be quite different. In particular, the
stationary irrotational flow of incompressible fluid is determined uniquely
by the boundary conditions. The rotational stationary flow may contain
stream lines, which do not cross the boundaries, and one cannot set the flow
problem, using only boundary conditions.

We consider the statement of the flow problem, starting from the simple case
of the arbitrary flow of the barotropic fluid, when the statement of the
problem is very simple. Imposing in series the constraints of
incompressibility and of stationary, we follow the evolution of the flow
problem statement.

In our investigation we use essentially the fact that dynamic equations for
the ideal barotropic fluid can be integrated \cite{R999}. Indefinite
functions appear in dynanic equations as a result of this integration. As a
rule, the investigation of integrals of differential equations is simpler
and more effective, than the investigation of differential equations
themselves, and we use this circumstance. Integration of hydrodynamic
equations is connected closely with a use of generalized stream function
(GSF) \cite{R04} and with GSF-technique which allows one to realize this
integration.

We obtain hydrodynamic equations for barotropic fluid from the variational
principle, which can be written in the form \cite{R04} 
\begin{equation}
\mathcal{A}_{\mathrm{E}}[\rho ,\mathbf{j},\mathbf{\xi },p]=\int \left\{ 
\frac{\mathbf{j}^{2}}{2\rho }-\rho E-p_{k}\left( j^{k}-\rho _{0}\left( 
\mathbf{\xi }\right) \frac{\partial J}{\partial \xi _{0,k}}\right) \right\}
dtd\mathbf{x}  \label{h1.1}
\end{equation}
where $j^{k}=\left\{ j^{0},\mathbf{j}\right\} =\left\{ \rho ,\rho \mathbf{v}%
\right\} $ is the 4-flux of the fluid, $\rho $ is the density and $\mathbf{v}
$ is its velocity. The quantity $E=E\left( \rho \right) $ is the fluid
internal energy per unit mass, which depends only on the density $\rho $.
The quantity $\rho _{0}=\rho _{0}\left( \mathbf{\xi }\right) $ is a given
weight function of $\mathbf{\xi }$. Variables $\mathbf{\xi }=\left\{ \xi
_{1},\xi _{2},\xi _{3}\right\} $ are Lagrangian coordinates, labeling the
fluid particles. Usually the Lagrangian coordinates are considered to be
independent variables. Here they are considered to be dependent variables $%
\mathbf{\xi }=\mathbf{\xi }\left\{ t,\mathbf{x}\right\} $. Considering $%
\mathbf{\xi }$ as dependent variables, we shall refer to them as Clebsch
potentials. These potentials have been used by Clebsch for description of
the incompressible fluid \cite{C57,C59}. In (\ref{h1.1}) and in what follows
a summation over repeated Latin indices is produced $(0-3)$. All dependent
dynamic variables $j$, $\mathbf{\xi }$, $p$ are considered to be functions
of $x=\left\{ x^{0},\mathbf{x}\right\} =\left\{ t,\mathbf{x}\right\} $.

The quantities $\partial J/\partial \xi _{0,k}$, $k=0,1,2,3$ are derivatives
of the Jacobian 
\begin{equation}
J\equiv \frac{\partial (\xi _{0},\xi _{1},\xi _{2},\xi _{3})}{\partial
(x^{0},x^{1},x^{2},x^{3})}\equiv \det \left| \left| \xi _{i,k}\right|
\right| ,\qquad \xi _{i,k}\equiv \partial _{k}\xi _{i}\equiv \frac{\partial
\xi _{i}}{\partial x^{k}},\qquad i,k=0,1,2,3  \label{h1.2}
\end{equation}
with respect to variables $\xi _{0,k}\equiv \partial _{k}\xi _{0}$. Here $%
\xi =\{\xi _{0},\mathbf{\xi }\}=\{\xi _{0},\xi _{1},\xi _{2},\xi _{3}\}$ are
four scalars considered to be functions of $x=\{x^{0},\mathbf{x}\}=\{t,%
\mathbf{x}\}$, $\xi =\xi (x)$. The functions $\{\xi _{0},\xi _{1},\xi
_{2},\xi _{3}\}$ are supposed to be independent in the sense that $J\neq 0$.
It is useful to consider the Jacobian $J$ as 4-linear function of variables $%
\xi _{i,k}\equiv \partial _{k}\xi _{i}$, $i,k=0,1,2,3$. Then one can
introduce derivatives of $J$ with respect to $\xi _{i,k}$. The derivative $%
\partial J/\partial \xi _{i,k}$ appears as a result of a replacement of $\xi
_{i}$ by $x^{k}$ in the relation (\ref{h1.2}).

\begin{equation}
\frac{\partial J}{\partial \xi _{i,k}}\equiv \frac{\partial (\xi _{0},...\xi
_{i-1},x^{k},{\xi }_{i+1}{,...}\xi _{3})}{\partial (x^{0},x^{1},x^{2},x^{3})}%
,\qquad i,k=0,1,2,3  \label{h1.3}
\end{equation}

Variables $\mathbf{\xi }=\left\{ \xi _{1},\xi _{2},\xi _{3}\right\} $ are
spatial Lagrangian coordinates of the fluid particles, whereas $\xi _{0}$ is
the temporal Lagrangian coordinate. It is fictitious in the action (\ref
{h1.1}). The quantities $\partial J/\partial \xi _{i,k}$ are useful, because
they satisfy identically to the relations 
\begin{equation}
\partial _{k}\frac{\partial J}{\partial \xi _{i,k}}\equiv 0,\qquad \frac{%
\partial J}{\partial \xi _{k,i}}\xi _{l,i}\equiv J\delta _{l}^{k},\qquad 
\frac{\partial J}{\partial \xi _{i,k}}\xi _{i,l}\equiv J\delta
_{l}^{k},\qquad l,k=0,1,2,3  \label{h1.4}
\end{equation}
Identifying the fluid 4-flux $j^{k}$ with the quantity $\rho _{0}\left( 
\mathbf{\xi }\right) \partial J/\partial \xi _{0,k}$ 
\begin{equation}
j^{k}=\rho _{0}\left( \mathbf{\xi }\right) \frac{\partial J}{\partial \xi
_{0,k}},\qquad k=0,1,2,3,  \label{h1.5}
\end{equation}
we obtain from two first equations (\ref{h1.4}) that the 4-flux $j^{k}$
satisfies the continuity equation 
\begin{equation}
\partial _{k}j^{k}=\partial _{k}\left( \rho _{0}\left( \mathbf{\xi }\right) 
\frac{\partial J}{\partial \xi _{0,k}}\right) =\rho _{0}\left( \mathbf{\xi }%
\right) \partial _{k}\frac{\partial J}{\partial \xi _{0,k}}+\frac{\partial
\rho _{0}\left( \mathbf{\xi }\right) }{\partial \xi _{\alpha }}\xi _{\alpha
,k}\frac{\partial J}{\partial \xi _{0,k}}\equiv 0  \label{h1.6}
\end{equation}
Here and in what follows a summation over two repeated Greek indices is
produced $(1-3)$. It follows from the second identity (\ref{h1.4}) that the
quantities $\mathbf{\xi }$ are labels of the fluid particles, and their
substantial derivatives vanish 
\begin{equation}
\left( \frac{\partial J}{\partial \xi _{0,0}}\right) ^{-1}\frac{\partial J}{%
\partial \xi _{0,k}}\partial _{k}\xi _{\alpha }=\frac{j^{k}}{\rho }\partial
_{k}\xi _{\alpha }=\left( \partial _{0}\xi _{\alpha }+\mathbf{v\nabla }\xi
_{\alpha }\right) =0,\qquad \alpha =1,2,3  \label{h1.7}
\end{equation}
A use of designation (\ref{h1.5}) is very useful, and we have introduced
this designation in the variational principle (\ref{h1.1}) by means of the
Lagrange multipliers $p_{k}=p_{k}\left( x\right) $,\ $k=0,1,2,3$.

To obtain hydrodynamic equations we should vary the action (\ref{h1.1}) with
respect to variables $\xi _{k},j^{k},p_{k}$,\ \ $k=0,1,2,3$. The variable $%
\xi _{0}$ is fictitious, and a variation with respect to $\xi _{0}$ gives
identity.

Dynamic equations have the form 
\begin{equation}
\frac{\delta \mathcal{A}}{\delta \xi _{i}}=-\partial _{l}\left( \rho
_{0}\left( \mathbf{\xi }\right) p_{k}\frac{\partial ^{2}J}{\partial \xi
_{0,k}\partial \xi _{i,l}}\right) +\frac{\partial \rho _{0}\left( \mathbf{%
\xi }\right) }{\partial \xi _{i}}p_{k}\frac{\partial J}{\partial \xi _{0,k}}%
=0,\qquad i=0,1,2,3  \label{h1.8}
\end{equation}
As far as the variable $\xi _{0}$ is fictitious, dynamic equation (\ref{h1.8}%
) with $i=0$ is to be an identity in force of other dynamic equations.
Another dynamic equations have the form 
\begin{equation}
\frac{\delta \mathcal{A}}{\delta j^{\alpha }}=\frac{j^{\alpha }}{\rho }%
-p_{\alpha }=0,\qquad \alpha =1,2,3  \label{h1.9}
\end{equation}
\begin{equation}
\frac{\delta \mathcal{A}}{\delta \rho }=-\frac{\mathbf{j}^{2}}{2\rho ^{2}}-%
\frac{\partial \left( \rho E\right) }{\partial \rho }-p_{0}=0  \label{h1.10}
\end{equation}
\begin{equation}
\frac{\delta \mathcal{A}}{\delta p_{k}}=-j^{k}+\rho _{0}\left( \mathbf{\xi }%
\right) \frac{\partial J}{\partial \xi _{0,k}}=0,\qquad k=0,1,2,3
\label{h1.11}
\end{equation}

Let us transform (\ref{h1.8}), using identities 
\begin{equation}
\partial _{l}\frac{\partial ^{2}J}{\partial \xi _{0,k}\partial \xi _{i,l}}%
\equiv 0,\qquad \frac{\partial ^{2}J}{\partial \xi _{0,k}\partial \xi _{i,l}}%
\equiv J^{-1}\left( \frac{\partial J}{\partial \xi _{0,k}}\frac{\partial J}{%
\partial \xi _{i,l}}-\frac{\partial J}{\partial \xi _{0,l}}\frac{\partial J}{%
\partial \xi _{i,k}}\right) ,\qquad i,k,l=0,1,2,3,  \label{h1.12}
\end{equation}
By means of the first identity (\ref{h1.12}) the equations (\ref{h1.8}) can
be written in the form 
\begin{equation}
-\frac{\partial ^{2}J}{\partial \xi _{0,k}\partial \xi _{i,l}}\rho
_{0}\left( \mathbf{\xi }\right) \partial _{l}p_{k}-\frac{\partial ^{2}J}{%
\partial \xi _{0,k}\partial \xi _{i,l}}p_{k}\partial _{l}\rho _{0}\left( 
\mathbf{\xi }\right) +\frac{\partial \rho _{0}\left( \mathbf{\xi }\right) }{%
\partial \xi _{i}}p_{k}\frac{\partial J}{\partial \xi _{0,k}}=0,\qquad
i=0,1,2,3  \label{h1.13}
\end{equation}
Two last terms of (\ref{h1.13}) compensate each other. Indeed, using the
second identity (\ref{h1.12}) we rewrite two last terms of (\ref{h1.13}) in
the form 
\begin{equation}
-J^{-1}\left( \frac{\partial J}{\partial \xi _{0,k}}\frac{\partial J}{%
\partial \xi _{i,l}}-\frac{\partial J}{\partial \xi _{0,l}}\frac{\partial J}{%
\partial \xi _{i,k}}\right) p_{k}\frac{\partial \rho _{0}\left( \mathbf{\xi }%
\right) }{\partial \xi _{\beta }}\xi _{\beta ,l}+\frac{\partial \rho
_{0}\left( \mathbf{\xi }\right) }{\partial \xi _{i}}p_{k}\frac{\partial J}{%
\partial \xi _{0,k}},\qquad i=0,1,2,3  \label{h1.14}
\end{equation}
In (\ref{h1.14}) and in what follows a summation over two repeated Greek
indices is produced $(1-3)$. Using the second identity (\ref{h1.4}), the
expression (\ref{h1.14}) is transformed to the form 
\begin{equation}
-\frac{\partial J}{\partial \xi _{0,k}}p_{k}\frac{\partial \rho _{0}\left( 
\mathbf{\xi }\right) }{\partial \xi _{\beta }}\delta _{\beta }^{i}+\frac{%
\partial \rho _{0}\left( \mathbf{\xi }\right) }{\partial \xi _{i}}p_{k}\frac{%
\partial J}{\partial \xi _{0,k}}=0,\qquad i=0,1,2,3  \label{h1.15}
\end{equation}
where two terms are compensated for $i=\beta =1,2,3$. For $i=0$ the first
term of (\ref{h1.15}) vanishes because of the multiplier $\delta _{\beta
}^{i}$, whereas the second term vanishes because $\partial \rho _{0}\left( 
\mathbf{\xi }\right) /\partial \xi _{0}=0$.

Thus, two last terms of (\ref{h1.13}) vanish, and using the second identity (%
\ref{h1.12}), the equation (\ref{h1.13}) takes the form 
\begin{equation}
-J^{-1}\left( \frac{\partial J}{\partial \xi _{0,k}}\frac{\partial J}{%
\partial \xi _{i,l}}-\frac{\partial J}{\partial \xi _{0,l}}\frac{\partial J}{%
\partial \xi _{i,k}}\right) \rho _{0}\left( \mathbf{\xi }\right)
p_{k,l}=0,\qquad i=0,1,2,3  \label{h1.16}
\end{equation}

Let us convolve (\ref{h1.16}) with $\xi _{i,s}$. Using the last identity (%
\ref{h1.4}) and the equation (\ref{h1.11}), we obtain from (\ref{h1.16})

\begin{equation}
\frac{\partial J}{\partial \xi _{0,k}}\left( p_{k,s}-p_{s,k}\right)
=0,\qquad s=0,1,2,3,\qquad p_{k,s}\equiv \partial _{s}p_{k}  \label{h1.17}
\end{equation}

It follows from (\ref{h1.9}) - (\ref{h1.11}) that 
\begin{equation}
\frac{\partial J}{\partial \xi _{0,0}}=\frac{\rho }{\rho _{0}\left( \mathbf{%
\xi }\right) }\text{,}\qquad p_{0}=-\frac{\mathbf{v}^{2}}{2}-\frac{\partial
\left( \rho E\right) }{\partial \rho },\qquad p_{\alpha }=v^{\alpha }\text{,}%
\qquad \frac{\partial J}{\partial \xi _{0,\alpha }}=\frac{\rho v^{\alpha }}{%
\rho _{0}\left( \mathbf{\xi }\right) },\qquad \alpha =1,2,3  \label{h1.18}
\end{equation}
Substituting (\ref{h1.18}) in (\ref{h1.17}), we obtain after transformations
for $s=\beta =1,2,3$%
\begin{equation}
v^{\alpha }{}_{,0}+v^{\beta }v^{\alpha }{}_{,\beta }=-\partial _{\alpha }%
\frac{\partial \left( \rho E\right) }{\partial \rho }=-\frac{1}{\rho }%
\partial _{\alpha }\left( \rho ^{2}\frac{\partial E}{\partial \rho }\right)
,\qquad \beta =1,2,3  \label{h1.19}
\end{equation}
and for $s=0$%
\begin{equation}
v^{\beta }v^{\beta }{}_{,0}+v^{\beta }\partial _{\beta }\left( \frac{\mathbf{%
v}^{2}}{2}+\frac{\partial \left( \rho E\right) }{\partial \rho }\right) =0
\label{h1.20}
\end{equation}
Here comma before index $k$ means differentiation with respect to $x^{k}$.
It is easy to see that (\ref{h1.20}) is a result of convolution of (\ref
{h1.19}) with $v^{\alpha }$. It is connected with that the equation (\ref
{h1.20}) appeared as a result of variation with respect to fictitious
variable $\xi _{0}$.

Equations (\ref{h1.6}), (\ref{h1.19}) and (\ref{h1.7}) form the complete
system of dynamic equations, which consists of seven first order
differential equations for seven dependent variables $\rho ,\mathbf{v},%
\mathbf{\xi }$. This system may be written in the vector form 
\begin{equation}
\partial _{0}\rho +\mathbf{\nabla }\left( \rho \mathbf{v}\right) =0
\label{h1.21}
\end{equation}
\begin{equation}
\partial _{0}\mathbf{v}+\left( \mathbf{v\nabla }\right) \mathbf{v}=-\frac{%
\mathbf{\nabla }p}{\rho },\qquad p=\rho ^{2}\frac{\partial E}{\partial \rho }
\label{h1.22}
\end{equation}
\begin{equation}
\partial _{0}\mathbf{\xi }+\left( \mathbf{v\nabla }\right) \mathbf{\xi }=0
\label{h1.23}
\end{equation}
Four equations (\ref{h1.21}), (\ref{h1.22}) form a closed subsystem (Euler
equations) of dynamic equations. These equations can be solved independently
of dynamic equations (\ref{h1.23}), which describe labeling of the fluid
particles and the character of the fluid particle motion along its
trajectory.

Indeed, if three quantities $\xi _{1}\left( t,\mathbf{x}\right) $, $\xi
_{2}\left( t,\mathbf{x}\right) $, $\xi _{3}\left( t,\mathbf{x}\right) $ are
three independent solution of equations (\ref{h1.23}) known as Lin
constraints \cite{L63}. They are three independent integrals of the system
of ordinary dynamic equations 
\begin{equation}
\frac{d\mathbf{x}}{dt}=\mathbf{v}\left( t,x\right)  \label{h1.24}
\end{equation}
If three independent integrals $\xi _{1}\left( t,\mathbf{x}\right) $, $\xi
_{2}\left( t,\mathbf{x}\right) $, $\xi _{3}\left( t,\mathbf{x}\right) $ of
the system (\ref{h1.4}) are known, the world lines (trajectories) of the
fluid particles $\mathbf{x}=\mathbf{x}\left( t,\mathbf{\xi }_{\mathrm{in}%
}\right) $ are determined implicitly by the algebraic equations 
\begin{equation}
\xi _{\alpha }\left( t,\mathbf{x}\right) =\left( \mathbf{\xi }_{\mathrm{in}%
}\right) _{\alpha }=\text{const},\qquad \alpha =1,2,3  \label{h1.25}
\end{equation}
Three quantities $\mathbf{\xi }_{\mathrm{in}}=\mathbf{\xi }$ label the fluid
particles and their world lines.

\section{Integration of dynamic equations for barotropic fluid}

There exists another form of hydrodynamic equations. The fact is that the
equations (\ref{h1.17}) can be integrated. Note that equations (\ref{h1.17})
are linear partial differential equations for the variables $%
p_{k},\;\;k=0,1,2,3$. They can be solved exactly in the form 
\begin{equation}
p_{k}=g^{0}\left( \xi _{0}\right) \xi _{0,k}+g^{\alpha }\left( \mathbf{\xi }%
\right) \xi _{\alpha ,k},\qquad k=0,1,2,3  \label{h2.1}
\end{equation}
where $\xi _{0}$ ceases to be fictitious and becomes to be a new dynamic
variable. The quantities $g^{0}$ and $g^{\alpha }$, $\alpha =1,2,3$ are
indefinite functions of arguments $\xi _{0}$ and $\mathbf{\xi }=\left\{ \xi
_{1},\xi _{2},\xi _{3}\right\} $ respectively.

Substituting (\ref{h2.1}) in (\ref{h1.17}) and using identities (\ref{h1.4}%
), one can verify that (\ref{h2.1}) is a solution of equations (\ref{h1.17})
for any functions $g^{i}$,\ $i=0,1,2,3$. It means that expression (\ref{h2.1}%
) gives the general solution of (\ref{h1.17}). Taking into account that the
first term in rhs of (\ref{h2.1}) is a gradient of some quantity $\varphi $,
we may write (\ref{h2.1}) in the form 
\begin{equation}
p_{k}=\partial _{k}\varphi +g^{\alpha }\left( \mathbf{\xi }\right) \xi
_{\alpha ,k},\qquad k=0,1,2,3  \label{h2.2}
\end{equation}
The first equation (\ref{h1.18}) takes the form 
\begin{equation}
\rho =\rho _{0}\left( \mathbf{\xi }\right) \frac{\partial J}{\partial \xi
_{0,0}}\equiv \rho _{0}\left( \mathbf{\xi }\right) \frac{\partial \left( \xi
_{1},\xi _{2},\xi _{3}\right) }{\partial \left( x,y,z\right) }\equiv \rho
_{0}\left( \mathbf{\xi }\right) \frac{\partial \left( \xi _{1},\xi _{2},\xi
_{3}\right) }{\partial \left( x^{1},x^{2},x^{3}\right) }  \label{h2.3}
\end{equation}
It follows from (\ref{h2.2}) and (\ref{h1.18}) 
\begin{equation}
v^{\mu }=\partial _{\mu }\varphi +g^{\alpha }\left( \mathbf{\xi }\right) \xi
_{\alpha ,\mu },\qquad \mu =1,2,3  \label{h2.4}
\end{equation}
Then equations (\ref{h1.23}) are transformed to the form 
\begin{equation}
\partial _{0}\xi _{\mu }+\left( \mathbf{\nabla }\varphi +g^{\alpha }\left( 
\mathbf{\xi }\right) \mathbf{\nabla }\xi _{\alpha }\right) \mathbf{\nabla }%
\xi _{\mu }=0,\qquad \mu =1,2,3  \label{h2.5}
\end{equation}
Let us set $k=0$ in (\ref{h2.2}). Eliminating $\xi _{\alpha ,0}$ by means of
(\ref{h2.5}), we obtain 
\begin{equation}
\partial _{0}\varphi -g^{\alpha }\left( \mathbf{\xi }\right) \left( \mathbf{%
\nabla }\varphi +g^{\beta }\left( \mathbf{\xi }\right) \mathbf{\nabla }\xi
_{\beta }\right) \mathbf{\nabla }\xi _{\alpha }+\frac{1}{2}\left( \mathbf{%
\nabla }\varphi +g^{\alpha }\left( \mathbf{\xi }\right) \mathbf{\nabla }\xi
_{\alpha }\right) ^{2}+P=0  \label{h2.6}
\end{equation}
\[
P=\left[ \frac{\partial \left( \rho E\right) }{\partial \rho }\right] _{\rho
=\rho _{0}\left( \mathbf{\xi }\right) \partial J/\partial \xi _{0,0}} 
\]
Equation (\ref{h2.6}) can be written in the form 
\begin{equation}
\partial _{0}\varphi +\frac{1}{2}\left( \mathbf{\nabla }\varphi \right) ^{2}-%
\frac{1}{2}g^{\alpha }\left( \mathbf{\xi }\right) g^{\beta }\left( \mathbf{%
\xi }\right) \mathbf{\nabla }\xi _{\beta }\mathbf{\nabla }\xi _{\alpha }+%
\left[ \frac{\partial \left( \rho E\right) }{\partial \rho }\right] _{\rho
=\rho _{0}\left( \mathbf{\xi }\right) \partial J/\partial \xi _{0,0}}=0
\label{h2.7}
\end{equation}
which allows one to interpret the variable $\varphi $. On one hand, the
variable $\varphi $ is a function of the temporal Lagrange coordinate $\xi
_{0}$. On the other hand, in the case, when $g^{\alpha }=0$ and $\varphi $
is the velocity potential, the equation (\ref{h2.7}) may be considered to be
the Hamilton--Jacobi equation with the Hamilton function 
\[
H\left( \mathbf{x,p}\right) =\frac{1}{2}\mathbf{p}^{2}+U\left( t,\mathbf{x}%
\right) ,\qquad U\left( t,\mathbf{x}\right) =\left[ \frac{\partial \left(
\rho E\right) }{\partial \rho }\right] _{\rho =\rho _{0}\left( \mathbf{\xi }%
\right) \partial J/\partial \xi _{0,0}} 
\]
In this case the Clebsch potential $\varphi $ may be regarded as the action
variable.

Thus, we have the system of four equations (\ref{h2.5}), (\ref{h2.6}) for
four dependent variables $\mathbf{\xi }$, $\varphi $. If solution of this
system (\ref{h2.5}), (\ref{h2.6}) has been obtained, the variables $\rho $, $%
\mathbf{v}$, are expressed via this solution by means of relations (\ref
{h2.3}), (\ref{h2.4}).

If we are interested in determination of the fluid flow, i.e. in
determination of variables $\rho $, $\mathbf{v}$ as functions of variables $%
t,\mathbf{x}$, we must solve either four Euler equations (\ref{h1.21}), (\ref
{h1.22}) with proper initial and boundary conditions, or four equations (\ref
{h2.5}), (\ref{h2.6}) with properly given functions $g^{\alpha }$, $\alpha
=1,2,3$ and properly given initial and boundary conditions for variables $%
\mathbf{\xi }$, $\varphi $.

Before comparative analyses of the two different systems of dynamic
equations we consider transition to the case of the incompressible fluid. To
pass to the incompressible fluid, we consider the slightly compressible
fluid with the internal energy of the form 
\begin{equation}
E\left( \rho \right) =E_{0}\left( \frac{\rho }{\rho _{1}}\right)
^{1/\varepsilon },\qquad E_{0},\rho _{1}=\text{const, \qquad }\varepsilon
\ll 1  \label{h2.8}
\end{equation}
The incompressible fluid appears in the limit $\varepsilon \rightarrow 0$.

Let us substitute (\ref{h2.8}) in (\ref{h2.6}) and resolve the obtained
relation with respect to the term, containing the constant $E_{0}$. We
obtain 
\begin{equation}
\left( \frac{1+\varepsilon }{\varepsilon }E_{0}\right) ^{\varepsilon }\frac{%
\rho }{\rho _{1}}=\left| \partial _{0}\varphi +\frac{1}{2}\left( \mathbf{%
\nabla }\varphi \right) ^{2}-\frac{1}{2}g^{\alpha }\left( \mathbf{\xi }%
\right) g^{\beta }\left( \mathbf{\xi }\right) \mathbf{\nabla }\xi _{\beta }%
\mathbf{\nabla }\xi _{\alpha }\right| ^{\varepsilon }  \label{h2.9}
\end{equation}
In the limit $\varepsilon \rightarrow 0$ the equations (\ref{h2.9}) and (\ref
{h2.3}) turn respectively into 
\begin{equation}
\rho =\rho _{1}=\text{const},\qquad \rho _{0}\left( \mathbf{\xi }\right) 
\frac{\partial \left( \xi _{1},\xi _{2},\xi _{3}\right) }{\partial \left(
x^{1},x^{2},x^{3}\right) }=\rho _{1}=\text{const}  \label{h2.10}
\end{equation}
Other dynamic equations of the system (\ref{h2.3}) - (\ref{h2.6}) do not
depend on $\rho $.

Conventional procedure of passage to the incompressible fluid in (\ref{h1.21}%
), (\ref{h1.22}) is an addition of the constraint 
\begin{equation}
\rho =\rho _{1}=\text{const}  \label{h2.11}
\end{equation}
to the Eulerian equations and elimination of connection between the density $%
\rho $ and the pressure $p$. As a result the pressure $p$ in (\ref{h1.22})
appears to be indefinite.

Taking into account (\ref{h2.11}), the Euler system of hydrodynamic
equations for the incompressible fluid takes the form 
\begin{equation}
\mathbf{\nabla v}=0,\qquad \partial _{0}\mathbf{v}+\left( \mathbf{v\nabla }%
\right) \mathbf{v}=-\frac{\mathbf{\nabla }p}{\rho _{0}},\qquad \rho _{0}=%
\text{const}  \label{h2.12}
\end{equation}
The equations (\ref{h2.3}) - (\ref{h2.6}) have the form 
\begin{equation}
\frac{\partial J}{\partial \xi _{0,0}}\equiv \frac{\partial \left( \xi
_{1},\xi _{2},\xi _{3}\right) }{\partial \left( x^{1},x^{2},x^{3}\right) }=%
\frac{\rho _{1}}{\rho _{0}\left( \mathbf{\xi }\right) },\qquad \rho _{1}=%
\text{const}  \label{h2.14a}
\end{equation}
\begin{equation}
\partial _{0}\xi _{\mu }+\left( \mathbf{\nabla }\varphi +g^{\alpha }\left( 
\mathbf{\xi }\right) \mathbf{\nabla }\xi _{\alpha }\right) \mathbf{\nabla }%
\xi _{\mu }=0,\qquad \mu =1,2,3  \label{h2.14}
\end{equation}
\begin{equation}
\mathbf{v}=\mathbf{\nabla }\varphi +g^{\alpha }\left( \mathbf{\xi }\right) 
\mathbf{\nabla }\xi _{\alpha }  \label{h2.15}
\end{equation}
In the system of integrated equations the condition 
\begin{equation}
\mathbf{\nabla v}=\mathbf{\nabla }^{2}\varphi +\mathbf{\nabla }\left(
g^{\alpha }\left( \mathbf{\xi }\right) \mathbf{\nabla }\xi _{\alpha }\right)
=0  \label{h2.16}
\end{equation}
takes place also, but it is not an independent relation. It is a corollary
of dynamic equations (\ref{h2.14a}), (\ref{h2.14}), (\ref{h2.15}).

Indeed, resolving (\ref{h2.14}) with respect to $v^{\alpha }=\partial
_{\alpha }\varphi +g^{\beta }\left( \mathbf{\xi }\right) \xi _{\beta ,\alpha
}$, we obtain in accordance with (\ref{h1.4}) 
\begin{equation}
v^{\alpha }=\partial _{\alpha }\varphi +g^{\beta }\left( \mathbf{\xi }%
\right) \xi _{\beta ,\alpha }=\left( \frac{\partial J}{\partial \xi _{0,0}}%
\right) ^{-1}\frac{\partial J}{\partial \xi _{0,\alpha }},\qquad \alpha
=1,2,3  \label{h2.17}
\end{equation}
Then (\ref{h2.17}) satisfies the relation (\ref{h2.16}), as it follows from (%
\ref{h1.6}) and (\ref{h2.14a}).

Thus, in the case of incompressible fluid we have three evolutional dynamic
equations, containing temporal derivatives, and one dynamic equation, which
does not contain temporal derivative (the first equation (\ref{h2.12}) and
the equation (\ref{h2.14a})). This equation is a constraint, imposed on the
state of incompressible fluid.

\section{Cauchy problem for barotropic fluid flow in \newline
infinite volume}

To obtain an unique solution for the barotropic fluid flow in the infinite
volume, one should give initial state $\rho ,\mathbf{v}$ of the fluid at the
time moment $t=0$. 
\begin{equation}
\rho \left( 0,\mathbf{x}\right) =\rho _{\mathrm{in}}\left( \mathbf{x}\right)
,\qquad \mathbf{v}\left( 0,\mathbf{x}\right) =\mathbf{v}_{\mathrm{in}}\left( 
\mathbf{x}\right)  \label{b3.1}
\end{equation}
Evolution of the fluid state $\rho ,\mathbf{v}$ is determined by evolutional
dynamic equations (\ref{h2.12}).

In the case of the integrated system (\ref{h2.3}) - (\ref{h2.6}) the initial
conditions (\ref{b3.1}) are to be given, but these conditions are not
sufficient for determination of unique solution of equations (\ref{h2.3}) - (%
\ref{h2.6}). One needs to give initial values for the Clebsch potentials $%
\varphi $, $\mathbf{\xi }$. We choose the simplest initial conditions for
the quantities $\varphi $, $\mathbf{\xi }$ 
\begin{equation}
\varphi \left( 0,\mathbf{x}\right) =\varphi _{\mathrm{in}}\left( \mathbf{x}%
\right) =0,\qquad \mathbf{\xi }\left( 0,\mathbf{x}\right) =\mathbf{\xi }_{%
\mathrm{in}}\left( \mathbf{x}\right) =\mathbf{x}  \label{b3.2}
\end{equation}

Substituting (\ref{b3.1}) and (\ref{b3.2}) in (\ref{h2.4}) we obtain 
\begin{equation}
\mathbf{g}\left( \mathbf{x}\right) =\mathbf{v}_{\mathrm{in}}\left( \mathbf{x}%
\right) ,\qquad \mathbf{g}\left( \mathbf{x}\right) =\left\{ g^{1}\left( 
\mathbf{x}\right) ,g^{2}\left( \mathbf{x}\right) ,g^{3}\left( \mathbf{x}%
\right) \right\}  \label{b3.3}
\end{equation}
Substituting (\ref{b3.1}) and (\ref{b3.2}) in dynamic equations (\ref{h2.3})
- (\ref{h2.7}), we obtain 
\begin{equation}
\rho _{0}\left( \mathbf{\xi }\right) =\rho _{\mathrm{in}}\left( \mathbf{x}%
\right) \left( \frac{\partial \left( \xi _{\mathrm{in}1},\xi _{\mathrm{in}%
2},\xi _{\mathrm{in}3}\right) }{\partial \left( x,y,z\right) }\right)
^{-1}=\rho _{\mathrm{in}}\left( \mathbf{x}\right) =\rho _{\mathrm{in}}\left( 
\mathbf{\xi }\right)  \label{b3.4}
\end{equation}
\begin{equation}
\partial _{0}\xi _{\mu }+\left( \mathbf{\nabla }\varphi +v_{\mathrm{in}%
}^{\alpha }\left( \mathbf{\xi }\right) \mathbf{\nabla }\xi _{\alpha }\right) 
\mathbf{\nabla }\xi _{\mu }=0,\qquad \mu =1,2,3  \label{b3.5}
\end{equation}
\begin{equation}
\partial _{0}\varphi +\frac{1}{2}\left( \mathbf{\nabla }\varphi \right) ^{2}-%
\frac{1}{2}v_{\mathrm{in}}^{\alpha }\left( \mathbf{\xi }\right) v_{\mathrm{in%
}}^{\beta }\left( \mathbf{\xi }\right) \mathbf{\nabla }\xi _{\beta }\mathbf{%
\nabla }\xi _{\alpha }+\left[ \frac{\partial \left( \rho E\right) }{\partial
\rho }\right] _{\rho =\rho _{0}\left( \mathbf{\xi }\right) \partial
J/\partial \xi _{0,0}}=0  \label{b3.6}
\end{equation}
Then relations (\ref{h2.3}), (\ref{h2.4}) take the form 
\begin{equation}
\rho =\rho _{\mathrm{in}}\left( \mathbf{\xi }\right) \frac{\partial \left(
\xi _{1},\xi _{2},\xi _{3}\right) }{\partial \left( x,y,z\right) },\qquad 
\mathbf{v}=\mathbf{\nabla }\varphi +v_{\mathrm{in}}^{\alpha }\left( \mathbf{%
\xi }\right) \mathbf{\nabla }\xi _{\alpha }  \label{b3.6a}
\end{equation}
where $\varphi $, $\mathbf{\xi }$ are solutions of (\ref{b3.6}), (\ref{b3.5}%
) with initial conditions (\ref{b3.2})

Choice of initial conditions for the Clebsch potentials $\varphi $, $\mathbf{%
\xi }$ in the form (\ref{b3.3}) is unessential. Variables $\mathbf{\xi }$
label the fluid particles, and one can use any single-valued method of
labeling. It means that equations (\ref{b3.5}), (\ref{b3.6}) are invariant
with respect to relabeling transformation. 
\begin{equation}
\xi _{\alpha }\rightarrow \tilde{\xi}_{\alpha }=\tilde{\xi}_{\alpha }\left( 
\mathbf{\xi }\right) ,\qquad v_{\mathrm{in}}^{\alpha }\left( \mathbf{\xi }%
\right) \rightarrow \tilde{v}_{\mathrm{in}}^{\alpha }\left( \mathbf{\tilde{%
\xi}}\right) =\frac{\partial \xi _{\beta }}{\partial \tilde{\xi}_{\alpha }}%
v_{\mathrm{in}}^{\beta }\left( \mathbf{\xi }\right) ,\qquad \alpha =1,2,3
\label{b3.7}
\end{equation}
Choice of initial condition $\varphi _{\mathrm{in}}$ in the form (\ref{b3.2}%
) is also unessential. Let us choose the initial conditions (\ref{h2.3}), (%
\ref{h2.4}) in the general form 
\begin{equation}
\varphi \left( 0,\mathbf{x}\right) =\varphi _{\mathrm{in}}\left( \mathbf{x}%
\right) ,\qquad \mathbf{\xi }\left( 0,\mathbf{x}\right) =\mathbf{\xi }_{%
\mathrm{in}}\left( \mathbf{x}\right) =\left\{ \xi _{\mathrm{in}1}\left( 
\mathbf{x}\right) ,\xi _{\mathrm{in}2}\left( \mathbf{x}\right) ,\xi _{%
\mathrm{in}3}\left( \mathbf{x}\right) \right\}   \label{b3.8}
\end{equation}
Substituting (\ref{b3.1}) and (\ref{b3.8}) in (\ref{h2.3}) and (\ref{h2.4}),
we obtain 
\begin{equation}
\rho _{\mathrm{in}}\left( \mathbf{x}\right) =\rho _{0}\left( \mathbf{\xi }_{%
\mathrm{in}}\left( \mathbf{x}\right) \right) D_{\mathrm{in}}\left( \mathbf{x}%
\right) ,\qquad g^{\alpha }\left( \mathbf{x}\right) =\left( v_{\mathrm{in}%
}^{\mu }\left( \mathbf{x}\right) -\partial _{\mu }\varphi _{\mathrm{in}%
}\left( \mathbf{x}\right) \right) \frac{\partial D_{\mathrm{in}}\left( 
\mathbf{x}\right) }{\partial \zeta _{\alpha ,\mu }},  \label{b3.9}
\end{equation}
\begin{equation}
D_{\mathrm{in}}\left( \mathbf{x}\right) \equiv \det \left| \left| \zeta
_{\alpha ,\beta }\right| \right| \equiv \frac{\partial \left( \xi _{\mathrm{%
in}1},\xi _{\mathrm{in}2},\xi _{\mathrm{in}3}\right) }{\partial \left(
x^{1},x^{2},x^{3}\right) },\qquad \zeta _{\alpha ,\beta }\equiv \frac{%
\partial \xi _{\mathrm{in}\alpha }}{\partial x^{\beta }},\qquad \alpha
,\beta =1,2,3  \label{b3.10}
\end{equation}
Then we have instead of (\ref{b3.6a}) 
\begin{equation}
\rho =\left[ \frac{\rho _{\mathrm{in}}\left( \mathbf{x}\right) }{D_{\mathrm{%
in}}\left( \mathbf{x}\right) }\right] _{\mathbf{x}=\mathbf{\xi }}\frac{%
\partial \left( \xi _{1},\xi _{2},\xi _{3}\right) }{\partial \left(
x^{1},x^{2},x^{3}\right) },  \label{b3.11}
\end{equation}
\begin{equation}
\mathbf{v}=\mathbf{\nabla }\varphi +\left( v_{\mathrm{in}}^{\mu }\left( 
\mathbf{\xi }\right) -\partial _{\mu }\varphi _{\mathrm{in}}\left( \mathbf{%
\xi }\right) \right) \left[ \frac{\partial D_{\mathrm{in}}}{\partial \zeta
_{\alpha ,\mu }}\left( \mathbf{x}\right) \right] _{\mathbf{x}=\mathbf{\xi }}%
\mathbf{\nabla }\xi _{\alpha }  \label{b3.11a}
\end{equation}
where $\varphi $, $\mathbf{\xi }$ are solutions of equations (\ref{h2.3}), (%
\ref{h2.7}) with the initial conditions (\ref{b3.8}),(\ref{b3.1}). These
equations have the form 
\begin{equation}
\partial _{0}\xi _{\mu }+\left( \mathbf{\nabla }\varphi +\left( v_{\mathrm{in%
}}^{\nu }\left( \mathbf{\xi }\right) -\partial _{\nu }\varphi _{\mathrm{in}%
}\left( \mathbf{\xi }\right) \right) \left[ \frac{\partial D_{\mathrm{in}}}{%
\partial \zeta _{\alpha ,\nu }}\left( \mathbf{x}\right) \right] _{\mathbf{x}=%
\mathbf{\xi }}\mathbf{\nabla }\xi _{\alpha }\right) \mathbf{\nabla }\xi
_{\mu }=0,\qquad \mu =1,2,3  \label{b3.12}
\end{equation}
\begin{eqnarray}
\partial _{0}\varphi  &=&-\frac{1}{2}\left( \mathbf{\nabla }\varphi \right)
^{2}-\left[ \frac{\partial \left( \rho E\right) }{\partial \rho }\right]
_{\rho =\rho _{0}\left( \mathbf{\xi }\right) \partial J/\partial \xi _{0,0}}
\nonumber \\
&&+\frac{1}{2}\left( v_{\mathrm{in}}^{\mu }\left( \mathbf{\xi }\right)
-\partial _{\mu }\varphi _{\mathrm{in}}\left( \mathbf{\xi }\right) \right) %
\left[ \frac{\partial D_{\mathrm{in}}}{\partial \zeta _{\alpha ,\mu }}\left( 
\mathbf{x}\right) \right] _{\mathbf{x}=\mathbf{\xi }}  \nonumber \\
&&\times \left( v_{\mathrm{in}}^{\nu }\left( \mathbf{\xi }\right) -\partial
_{\nu }\varphi _{\mathrm{in}}\left( \mathbf{\xi }\right) \right) \left[ 
\frac{\partial D_{\mathrm{in}}}{\partial \zeta _{\beta ,\nu }}\left( \mathbf{%
x}\right) \right] _{\mathbf{x}=\mathbf{\xi }}\mathbf{\nabla }\xi _{\alpha }%
\mathbf{\nabla }\xi _{\beta }  \label{b3.14}
\end{eqnarray}
where 
\begin{equation}
\rho _{0}\left( \mathbf{\xi }\right) =\left[ \frac{\rho _{\mathrm{in}}\left( 
\mathbf{x}\right) }{D_{\mathrm{in}}\left( \mathbf{x}\right) }\right] _{%
\mathbf{x}=\mathbf{\xi }}  \label{b3.15}
\end{equation}
Equations (\ref{b3.12}), (\ref{b3.14}) should be solved at the initial
conditions (\ref{b3.8}). Instead we can also solve equations (\ref{b3.5}), (%
\ref{b3.6}) at the initial conditions (\ref{b3.2}).

We see that the integrated dynamic equations (\ref{h2.5}), (\ref{h2.7}) (or
in expanded form (\ref{b3.12}), (\ref{b3.14}), (\ref{b3.15})) contain full
information on the fluid flow. Initial conditions (\ref{b3.8}), which are
necessary for determination of the unique solution of dynamic equations (\ref
{h2.5}), (\ref{h2.7}), concern only physically unessential information on
the fluid particles labeling and separation of the velocity into potential
and vortical components.

If we consider the Lagrangian coordinates $\mathbf{\xi }$ as independent
variables the dynamic equations (\ref{b3.5}), (\ref{b3.6}) and (\ref{b3.6a})
are reduced to the form 
\begin{equation}
\varphi ^{,0}-\frac{1}{2}\left( \varphi ^{,\alpha }+v_{\mathrm{in}}^{\alpha
}\left( \mathbf{\xi }\right) \right) \left( \varphi ^{,\alpha }+v_{\mathrm{in%
}}^{\alpha }\left( \mathbf{\xi }\right) \right) X^{-2}\frac{\partial X}{%
\partial x^{\mu ,\alpha }}\frac{\partial X}{\partial x^{\mu ,\alpha }}%
+P\left( \rho \right) =0,  \label{b3.16}
\end{equation}
\begin{equation}
x^{\beta ,0}=\left( \frac{\partial \varphi }{\partial \xi _{\alpha }}+v_{%
\mathrm{in}}^{\alpha }\left( \mathbf{\xi }\right) \right) X^{-1}\frac{%
\partial X}{\partial x^{\beta ,\alpha }},\qquad \beta =1,2,3  \label{b3.17}
\end{equation}
\begin{equation}
P\left( \rho \right) =\left[ \frac{\partial \left( \rho E\left( \rho \right)
\right) }{\partial \rho }\right] _{\rho =X^{-1}\rho _{\mathrm{in}}\left( 
\mathbf{\xi }\right) }  \label{b3.17a}
\end{equation}
where $\mathbf{x}=\left\{ x^{\alpha }\left( t,\mathbf{\xi }\right) \right\}
, $\ $\alpha =1,2,3$, and $\varphi =\varphi \left( t,\mathbf{\xi }\right) $.
Jacobian 
\begin{equation}
X=\frac{\partial \left( x^{1},x^{2},x^{3}\right) }{\partial \left( \xi
_{1},\xi _{2},\xi _{3}\right) }=\det \left| \left| x^{\alpha ,\beta }\right|
\right| ,\qquad x^{\alpha ,\beta }\equiv \frac{\partial x^{\alpha }}{%
\partial \xi _{\beta }},\qquad \alpha ,\beta =1,2,3  \label{b3.18}
\end{equation}
is considered to be a function of variables $x^{\alpha ,\beta }$. The
quantities of the type $u^{,0}$ mean the time derivative of $u$ with
constant $\mathbf{\xi }$ 
\begin{equation}
x^{\beta ,0}\equiv \frac{dx^{\beta }}{dt}=\frac{\partial \left( x^{\beta
},\xi _{1},\xi _{2},\xi _{3}\right) }{\partial \left( t,\xi _{1},\xi
_{2},\xi _{3}\right) },\qquad \varphi ^{,0}\equiv \frac{d\varphi }{dt}=\frac{%
\partial \left( \varphi ,\xi _{1},\xi _{2},\xi _{3}\right) }{\partial \left(
t,\xi _{1},\xi _{2},\xi _{3}\right) }  \label{b3.19}
\end{equation}
Hydrodynamic equations (\ref{b3.16}), (\ref{b3.17}) are rather bulky, but
they contain arbitrary initial conditions as functions of independent
variables $\mathbf{\xi }$.

\section{Cauchy problem for the incompressible fluid flow in infinite volume}

The main difference between the barotropic and incompressible fluids
consists in the constraint imposed on the state of the incompressible fluid
by the first equation (\ref{h2.12}). This condition does not contain
temporal derivative and it is to be satisfied at the initial moment $t=0$%
\begin{equation}
\mathbf{\nabla v}_{\mathrm{in}}\left( \mathbf{x}\right) =0  \label{h4.1}
\end{equation}
It means that the initial state of the incompressible fluid $\mathbf{v}_{%
\mathrm{in}}$ cannot be given arbitrarily. But the main property of initial
state is the possibility of giving it arbitrarily. To conserve this
property, we are forced to redefine the concept of initial state of the
incompressible fluid. Let us consider the generalized stream function (GSF) $%
\left\{ \psi _{2},\psi _{3}\right\} $ \cite{R04} to be the quantity
describing the state of the incompressible fluid. The velocity $\mathbf{v}$,
defined via GSF by the relation 
\begin{equation}
v^{\mu }=\frac{\partial \left( x^{\mu },\psi _{2},\psi _{3}\right) }{%
\partial \left( x^{1},x^{2},x^{3}\right) },\qquad \mu =1,2,3,  \label{h4.2}
\end{equation}
satisfies the first equation (\ref{h2.12}) for any choice of functions $%
\left\{ \psi _{2},\psi _{3}\right\} $. Considering GSF as a state of the
incompressible fluid, we use two quantities $\left\{ \psi _{2},\psi
_{3}\right\} $ instead of three components of velocity, but these quantities 
$\left\{ \psi _{2},\psi _{3}\right\} $ may be given arbitrarily at the
initial time $t=0$. It is very useful and important.

On the other hand, the variables $j^{0}=\rho $ and $\mathbf{j}=\rho \mathbf{v%
}$ are described by relations (\ref{h1.11}). Choosing $\rho _{0}\left( 
\mathbf{\xi }\right) \mathbf{=}\rho _{1}=$const, we obtain in the case of
the incompressible fluid from (\ref{h1.11})

\begin{equation}
\rho =\rho _{1}\frac{\partial \left( \xi _{1},\xi _{2},\xi _{3}\right) }{%
\partial \left( x^{1},x^{2},x^{3}\right) }=\rho _{1}=\text{const},\qquad
j^{\mu }=\rho _{1}v^{\mu }=\rho _{1}\frac{\partial \left( x^{\mu },\xi
_{1},\xi _{2},\xi _{3}\right) }{\partial \left(
x^{0},x^{1},x^{2},x^{3}\right) },\qquad \mu =1,2,3  \label{h4.3}
\end{equation}
The velocity $\mathbf{v}$, defined by the second equation (\ref{h4.3}),
satisfies the first equation (\ref{h2.12}) identically, because 
\begin{equation}
\frac{\partial \left( \xi _{1},\xi _{2},\xi _{3}\right) }{\partial \left(
x^{1},x^{2},x^{3}\right) }=1  \label{h4.4}
\end{equation}

Let us choose initial conditions in the form 
\begin{equation}
\mathbf{\xi }\left( 0,\mathbf{x}\right) =\mathbf{\xi }_{\mathrm{in}}\left( 
\mathbf{x}\right) =\mathbf{x,\qquad }\varphi \left( 0,\mathbf{x}\right)
=\varphi _{\mathrm{in}}\left( \mathbf{x}\right) =0  \label{h4.5}
\end{equation}
\begin{equation}
v^{\mu }(0,\mathbf{x})=v_{\mathrm{in}}^{\mu }(\mathbf{x})=\frac{\partial
\left( x^{\mu },\psi _{\mathrm{in}2}\left( \mathbf{x}\right) ,\psi _{\mathrm{%
in}3}\left( \mathbf{x}\right) \right) }{\partial \left(
x^{1},x^{2},x^{3}\right) },\qquad \mu =1,2,3  \label{h4.7}
\end{equation}
where $\psi _{\mathrm{in}2}$, $\psi _{\mathrm{in}3}$ are given functions of $%
\mathbf{x}$ (initial values of GSF). Then it follows from (\ref{h2.15})
written at $t=0$%
\begin{equation}
g^{\mu }\left( \mathbf{x}\right) =\frac{\partial \left( x^{\mu },\ \psi _{%
\mathrm{in}2}\left( \mathbf{x}\right) ,\psi _{\mathrm{in}3}\left( \mathbf{x}%
\right) \right) }{\partial \left( x^{1},x^{2},x^{3}\right) },\qquad \mu
=1,2,3  \label{h4.8}
\end{equation}
Substituting (\ref{h4.8}) in dynamic equations (\ref{h2.14}) and setting $%
\rho _{1}/\rho _{0}\left( \mathbf{x}\right) =$const, we obtain three dynamic
equations for variables $\mathbf{\xi }$%
\begin{equation}
\partial _{0}\xi _{\alpha }+\left( \varphi _{,\nu }+\frac{\partial \left(
\xi _{\mu },\psi _{\mathrm{in}2}\left( \mathbf{\xi }\right) ,\psi _{\mathrm{%
in}3}\left( \mathbf{\xi }\right) \right) }{\partial \left( \xi _{1},\xi
_{2},\xi _{3}\right) }\xi _{\mu ,\nu }\right) \xi _{\alpha ,\nu }=0,\qquad
\alpha =1,2,3  \label{h4.20}
\end{equation}
and a constraint (\ref{h4.4}), imposed on the values of the quantities $%
\mathbf{\xi }$.

Four equations (\ref{h4.20}), (\ref{h4.4}) form a system of dynamic
equations for four dynamic variables $\mathbf{\xi ,}\varphi $. Equation (\ref
{h4.4}) may be replaced by the equation (\ref{h2.16}), which after
substitution of (\ref{h4.8}) takes the form 
\begin{equation}
\varphi _{,\nu \nu }+\partial _{\nu }\left( \frac{\partial \left( \xi _{\mu
},\psi _{\mathrm{in}2}\left( \mathbf{\xi }\right) ,\psi _{\mathrm{in}%
3}\left( \mathbf{\xi }\right) \right) }{\partial \left( \xi _{1},\xi
_{2},\xi _{3}\right) }\xi _{\mu ,\nu }\right) =0  \label{h4.22}
\end{equation}
where $\psi _{\mathrm{in}2}\left( \mathbf{\xi }\right) $, $\psi _{\mathrm{in}%
3}\left( \mathbf{\xi }\right) $ are given functions of argument $\mathbf{\xi 
}$\textbf{. }We stress that the equations (\ref{h4.20}), (\ref{h4.22}) are
to be solved at initial conditions (\ref{h4.5}). May we set $\varphi =0$ in
the equation (\ref{h4.22})? In general, no. If we set $\varphi =0$, the
equation (\ref{h4.22}) turns into the equation 
\begin{equation}
\partial _{\nu }\left( \frac{\partial \left( \xi _{\mu },\psi _{\mathrm{in}%
2}\left( \mathbf{\xi }\right) ,\psi _{\mathrm{in}3}\left( \mathbf{\xi }%
\right) \right) }{\partial \left( \xi _{1},\xi _{2},\xi _{3}\right) }\xi
_{\mu ,\nu }\right) =0  \label{h4.23}
\end{equation}
which is valid at $t=0$, when $\mathbf{\xi }=\mathbf{x}$. If $\mathbf{\xi }%
\neq \mathbf{x}$, equation (\ref{h4.23}) is not valid, in general.

Thus, four equations (\ref{h4.20}), (\ref{h4.22}) form a system of dynamic
equations for four dynamic variables $\mathbf{\xi }$, $\varphi $. Dynamic
equations (\ref{h4.20}) are evolutional equations, describing evolution of $%
\mathbf{\xi }$ in the sense that they contain time derivatives of $\mathbf{%
\xi }$. If the state $\mathbf{\xi }\left( t,\mathbf{x}\right) $, $\varphi
\left( t,\mathbf{x}\right) $ of the fluid is given at the time $t$, the
dynamic equations (\ref{h4.20}) determine the quantities $\mathbf{\xi }%
\left( t+dt,\mathbf{x}\right) $ at the next time moment uniquely.

The\textbf{\ }equation (\ref{h4.22}) as well as the equation (\ref{h4.4}) is
not an evolutional equation, because it does not contain temporal
derivatives. The value $\varphi \left( t+dt,\mathbf{x}\right) $ is not
connected with the value $\varphi \left( t,\mathbf{x}\right) $ directly. One
can determine the unique solution of (\ref{h4.22}), if there is some
additional information about the variable $\varphi $ on the boundary $%
\partial V$ of the volume $V$, where the fluid flow is considered. Formally
it follows from the fact that the Poisson equation (\ref{h4.22}) has an
unique solution in the region $V$, provided a proper information is given on
the boundary $\partial V$ of the volume $V$.

In the case of the barotropic fluid the corresponding equation (\ref{b3.14})
for $\varphi $ is evolutional, and one does not need such an information. At
least at the point $\mathbf{x}$, which is far enough from the boundary $%
\partial V$, the values of $\varphi \left( t,\mathbf{x}\right) ,$\ \ $0<t$ $%
<T$ can be determined uniquely for a time $T$, which is necessary for
passage of the signal from the nearest point of boundary $\partial V$ to the
point $\mathbf{x}$. The speed of the signal is equal to the sound speed in
the fluid. In the incompressible fluid the sound speed is infinite, the time
interval $T=0$, and dynamic equation (\ref{h4.22}) is not evolutional. In
accordance with this fact the value $\varphi \left( t+dt,\mathbf{x}\right) $
depends on the boundary conditions at the time $t+dt$, but not on the values
of $\varphi \left( t,\mathbf{x}\right) $ and $\mathbf{\xi }\left( t,\mathbf{x%
}\right) $.

One can see two parts in dynamic equations (\ref{h4.20}): nonlocal potential
term $\varphi _{,\nu }\xi _{\alpha ,\nu }$ and local vortical term $v_{%
\mathrm{in}}^{\mu }\xi _{\mu ,\nu }\xi _{\alpha ,\nu }$. The vortical term
describes influence of the ''frozen'' vorticity on the fluid flow. The
potential term describes nonlocal interaction in the fluid connected with
the potential $\varphi $ and with the infinite speed of sound. In the
barotropic fluid the speed of sound is finite, and the potential term
describes a local interaction.

\section{Stationary flow of incompressible fluid.}

Let us imagine that setting the Cauchy problem, we choose the initial
conditions in such a way that the incompressible fluid flow appears to be
stationary, i.e. 
\begin{equation}
\partial _{0}\rho =0,\qquad \partial _{0}\mathbf{v}=0  \label{h5.1}
\end{equation}
In this case the stream lines are stationary. Let Clebsch potentials $\xi
_{2}$ and $\xi _{3}$ label these stream lines. The Clebsch potentials $\xi
_{2}$ and $\xi _{3}$ can be chosen independent of time. Dynamic equations (%
\ref{h2.14a}), (\ref{h2.14}), (\ref{h2.15}), (\ref{h2.17}) take the form 
\begin{equation}
\frac{\partial J}{\partial \xi _{0,0}}\equiv \rho _{1}\frac{\partial \left(
\xi _{1},\xi _{2},\xi _{3}\right) }{\partial \left( x^{1},x^{2},x^{3}\right) 
}=\rho _{1}=\text{const}  \label{h5.1a}
\end{equation}
\begin{equation}
v^{\mu }=\frac{\partial J}{\partial \xi _{0,\mu }}\equiv \frac{\partial
\left( x^{\mu },\xi _{1},\xi _{2},\xi _{3}\right) }{\partial \left(
x^{0},x^{1},x^{2},x^{3}\right) }=-\xi _{1,0}\frac{\partial \left( x^{\mu
},\xi _{2},\xi _{3}\right) }{\partial \left( x^{1},x^{2},x^{3}\right) }%
,\qquad \mu =1,2,3  \label{h5.2}
\end{equation}
\begin{equation}
\partial _{0}\xi _{1}+\left( \mathbf{v\nabla }\right) \xi _{1}=0,\qquad
\left( \mathbf{v\nabla }\right) \xi _{2}=0,\qquad \left( \mathbf{v\nabla }%
\right) \xi _{3}=0  \label{s5.3}
\end{equation}
\begin{equation}
\mathbf{v}=\mathbf{\nabla }\varphi +g^{\alpha }\left( \mathbf{\xi }\right) 
\mathbf{\nabla }\xi _{\alpha }  \label{s5.4}
\end{equation}

Let us write equations (\ref{h5.2}), (\ref{h5.1a}) respectively in the form 
\begin{equation}
\mathbf{v}=-\xi _{1,0}\left( \mathbf{\nabla }\xi _{2}\times \mathbf{\nabla }%
\xi _{3}\right) ,\qquad \mathbf{\nabla }\xi _{1}\left( \mathbf{\nabla }\xi
_{2}\times \mathbf{\nabla }\xi _{3}\right) =1  \label{s5.4a}
\end{equation}
Let us set $\xi _{1,0}=-1$. We obtain 
\begin{equation}
\mathbf{v}=\left( \mathbf{\nabla }\xi _{2}\times \mathbf{\nabla }\xi
_{3}\right) ,\qquad \xi _{1,0}=-1  \label{h5.2a}
\end{equation}

The first equation (\ref{s5.3}) takes the form 
\begin{equation}
\left( \mathbf{v\nabla }\right) \xi _{1}=1,  \label{s5.6}
\end{equation}
We integrate the second equation (\ref{s5.4a}), considering variables $\xi
_{2},\xi _{3}$ as given functions of argument $\mathbf{x}$. It can be
written in the form 
\begin{equation}
\frac{\partial \left( \xi _{1},\xi _{2},\xi _{3}\right) }{\partial \left(
x,y,z\right) }=\frac{\partial \left( \xi _{1},\xi _{2},\xi _{3}\right) }{%
\partial \left( s,\xi _{2},\xi _{3}\right) }D\left( \mathbf{x}\right)
=1,\qquad D\left( \mathbf{x}\right) =\frac{\partial \left( s,\xi _{2},\xi
_{3}\right) }{\partial \left( x,y,z\right) }  \label{s5.7}
\end{equation}
where $s$ is some function of $\mathbf{x}$, which is chosen in such a way,
that 
\begin{equation}
D\left( \mathbf{x}\right) =\frac{\partial \left( s,\xi _{2},\xi _{3}\right) 
}{\partial \left( x,y,z\right) }\neq 0  \label{s5.8}
\end{equation}
Let us resolve equations 
\begin{equation}
\xi _{2}=\xi _{2}\left( x,y,z\right) ,\qquad \xi _{3}=\xi _{3}\left(
x,y,z\right) ,\qquad s=s\left( x,y,z\right)  \label{s5.8a}
\end{equation}
in the form 
\begin{equation}
x=F_{1}\left( s,\xi _{2},\xi _{3}\right) ,\qquad y=F_{2}\left( s,\xi
_{2},\xi _{3}\right) ,\qquad z=F_{3}\left( s,\xi _{2},\xi _{3}\right)
\label{s5.8b}
\end{equation}
and calculate 
\begin{equation}
\Delta \left( s,\xi _{2},\xi _{3}\right) =D\left( F_{1}\left( s,\xi _{2},\xi
_{3}\right) ,F_{2}\left( s,\xi _{2},\xi _{3}\right) ,F_{3}\left( s,\xi
_{2},\xi _{3}\right) \right)  \label{s5.8c}
\end{equation}

The equation (\ref{s5.7}) can be written in the form 
\begin{equation}
\frac{\partial \left( \xi _{1},\xi _{2},\xi _{3}\right) }{\partial \left(
s,\xi _{2},\xi _{3}\right) }=\frac{1}{\Delta \left( s,\xi _{2},\xi
_{3}\right) }  \label{s5.9}
\end{equation}
It is integrated in the form 
\begin{equation}
\xi _{1}=\int \frac{ds}{\Delta \left( s,\xi _{2},\xi _{3}\right) }
\label{s5.10}
\end{equation}
where integration is produced at fixed $\xi _{2},\xi _{3}$.

We eliminate the variable $\varphi $ from the equation (\ref{s5.4}), taking
its curl. We obtain 
\begin{equation}
\mathbf{\nabla }\times \mathbf{v}=\frac{1}{2}\Omega ^{\alpha \beta }\left( 
\mathbf{\nabla }\xi _{\alpha }\times \mathbf{\nabla }\xi _{\beta }\right)
=\Omega ^{23}\left( \mathbf{\nabla }\xi _{2}\times \mathbf{\nabla }\xi
_{3}\right) +\Omega ^{31}\left( \mathbf{\nabla }\xi _{3}\times \mathbf{%
\nabla }\xi _{1}\right) +\Omega ^{12}\left( \mathbf{\nabla }\xi _{1}\times 
\mathbf{\nabla }\xi _{2}\right)  \label{s5.11}
\end{equation}
where 
\begin{equation}
\Omega ^{\alpha \beta }=\Omega ^{\alpha \beta }\left( \mathbf{\xi }\right)
\equiv \frac{\partial g^{a}}{\partial \xi _{\beta }}\left( \mathbf{\xi }%
\right) -\frac{\partial g^{\beta }}{\partial \xi _{\alpha }}\left( \mathbf{%
\xi }\right) =g^{a,\beta }\left( \mathbf{\xi }\right) -g^{\beta ,\alpha
}\left( \mathbf{\xi }\right)  \label{s5.12}
\end{equation}
The quantities $\Omega ^{\alpha \beta }$ describe the fluid flow vorticity
in the coordinates $\xi _{1},\xi _{2},\xi _{3}$, as it follows from the
relation (\ref{s5.11}), written in the form

\begin{equation}
\partial _{\mu }v^{\nu }\left( \mathbf{x}\right) -\partial _{\nu }v^{\mu
}\left( \mathbf{x}\right) =\frac{\partial \xi _{\alpha }}{\partial x^{\mu }}%
\frac{\partial \xi _{\beta }}{\partial x^{\nu }}\Omega ^{\alpha \beta
}\left( \mathbf{\xi }\right)  \label{s5.22}
\end{equation}
This relation may interpreted as a transformation of the tensor $\partial
_{\mu }v^{\nu }-\partial _{\nu }v^{\mu }$ from the coordinates $\mathbf{\xi }
$ to coordinates $\mathbf{x}$.

The variables $\xi _{2}$ and $\xi _{3}$ do not depend on $t$, whereas $\xi
_{1}$ is a linear function of $t$. Left hand side of (\ref{s5.11}) does not
depend on $t$. It means that the quantities $\Omega ^{\alpha \beta }$ do not
depend on $t$ also. They depend only on $\xi _{2},\xi _{3}$, but not on $\xi
_{1}$, because $\xi _{1}$ depends on $t$. Differentiating (\ref{s5.10}), we
obtain 
\begin{equation}
\mathbf{\nabla }\xi _{1}\times \mathbf{\nabla }\xi _{2}=\frac{\mathbf{\nabla 
}s\times \mathbf{\nabla }\xi _{2}}{\Delta \left( s,\xi _{2},\xi _{3}\right) }%
+\left( \mathbf{\nabla }\xi _{3}\times \mathbf{\nabla }\xi _{2}\right) \frac{%
\partial }{\partial \xi _{3}}\int \frac{ds}{\Delta \left( s,\xi _{2},\xi
_{3}\right) }  \label{s5.14}
\end{equation}
\begin{equation}
\mathbf{\nabla }\xi _{1}\times \mathbf{\nabla }\xi _{3}=\frac{\mathbf{\nabla 
}s\times \mathbf{\nabla }\xi _{3}}{\Delta \left( s,\xi _{2},\xi _{3}\right) }%
+\left( \mathbf{\nabla }\xi _{2}\times \mathbf{\nabla }\xi _{3}\right) \frac{%
\partial }{\partial \xi _{2}}\int \frac{ds}{\Delta \left( s,\xi _{2},\xi
_{3}\right) }  \label{s5.15}
\end{equation}

Substituting relations (\ref{s5.14}) and (\ref{s5.15}) in (\ref{s5.11}) and
using (\ref{h5.2a}), we obtain 
\begin{eqnarray}
\mathbf{\nabla }\times \mathbf{v} &=&\left( \Omega ^{23}-\Omega ^{31}\frac{%
\partial }{\partial \xi _{2}}\int \frac{ds}{\Delta \left( s,\xi _{2},\xi
_{3}\right) }-\Omega ^{12}\frac{\partial }{\partial \xi _{3}}\int \frac{ds}{%
\Delta \left( s,\xi _{2},\xi _{3}\right) }\right) \mathbf{v}  \nonumber \\
&&+\frac{\mathbf{\nabla }s\times \left( \Omega ^{12}\mathbf{\nabla }\xi
_{2}-\Omega ^{31}\mathbf{\nabla }\xi _{3}\right) }{\Delta \left( s,\xi
_{2},\xi _{3}\right) },\qquad \Omega ^{\alpha \beta }=\Omega ^{\alpha \beta
}\left( \xi _{2},\xi _{3}\right)  \label{s5.16}
\end{eqnarray}
Three independent equations (\ref{s5.16}) carry out the description of the
stationary flow of incompressible fluid in terms of the generalized stream
function (GSF) $\left\{ \xi _{2},\xi _{3}\right\} $. The velocity $\mathbf{v}
$ is expressed via GSF by means of the relation (\ref{h5.2a}). The variable $%
s$ is chosen in such a way, that the inequality (\ref{s5.8}) takes place.
For instance, one can set $s=x$. Then we obtain from (\ref{s5.8}) 
\[
\Delta =\frac{\partial \left( x,\xi _{2},\xi _{3}\right) }{\partial \left(
x,y,z\right) }=v^{1}\left( x,\xi _{2},\xi _{3}\right) 
\]
In this case the equations (\ref{s5.16}) has a singular point, when the
velocity component $v^{1}$ vanishes. At another choice of the variable $s$
the singular point does not appear, or it appears in other place. It means
that the possible singular point is a result of unsuccessful description,
which can be eliminated by a proper choice of the variable $s$.

The conventional statement of the stationary flow problem of the
incompressible fluid can be obtained from the dynamic equations (\ref{h2.12}%
) and (\ref{h5.1}). It has the form 
\begin{equation}
\mathbf{\nabla v}=0,\qquad \mathbf{\nabla }\times \left( \mathbf{v\nabla }%
\right) \mathbf{v}=0  \label{s5.17}
\end{equation}
The first equation (\ref{s5.17}) can be solved by introduction of GSF $%
\left\{ \xi _{2},\xi _{3}\right\} $. Then the velocity is expressed by means
of (\ref{h5.2a}), and we obtain the dynamic equations for the stationary
flow of the incompressible fluid 
\begin{equation}
\mathbf{\nabla }\times \left( \mathbf{v\nabla }\right) \mathbf{v}=0,\qquad 
\mathbf{v}=\left( \mathbf{\nabla }\xi _{2}\times \mathbf{\nabla }\xi
_{3}\right)  \label{s5.18}
\end{equation}

Let us compare two different statement of the problem of the stationary flow
of the incompressible fluid. We shall compare equations (\ref{s5.16}) and (%
\ref{s5.17}). At first, we consider equations (\ref{s5.17}). In the case of
the irrotational flow, when 
\begin{equation}
\mathbf{v}=\mathbf{\nabla }\varphi  \label{s5.19}
\end{equation}
the second equation (\ref{s5.17}) is satisfied identically. The first
equation (\ref{s5.17}) leads to the Laplace equation 
\begin{equation}
\mathbf{\nabla }^{2}\varphi =0  \label{s5.20}
\end{equation}
which has a unique solution, provided that the normal derivative $\left( 
\mathbf{n\nabla }\right) \varphi $ is given on the boundary $\partial V$ of
the volume $V$, where the fluid flow is considered. In other words, for
determination of an unique irrotational flow it is sufficient to give the
velocity $\mathbf{v}$ on the boundary $\partial V$. What information is
necessary for determination of the unique rotational flow? It is believed
that in the case of the rotational flow the same information is sufficient
as in the case of the irrotational flow. Why? Because in the case of the
nonstationary flow it is sufficient to give the initial velocity $\mathbf{v}%
_{\mathrm{in}}\left( \mathbf{x}\right) $ in the whole volume $V$ and the
velocity $\mathbf{v}_{\mathrm{b}}\left( t,,\mathbf{x}\right) ,$ $\mathbf{x}%
\in \partial V$ on the boundary $\partial V$. This information is the same
for both rotational and irrotational flow. In the stationary case the
determination of the rotational flow needs additional information as
compared with the irrotational flow. It is necessary to give the vorticity $%
\Omega ^{\alpha \beta }$ on each stream line. In reality, the vorticity is
given also in the case of the stationary irrotational flow. It is simply
equal to zero, and this information is perceived usually as an absence of
information.

Why is it necessary to give the vorticity in addition to the velocity in the
stationary case? Why may we not give the vorticity in the nonstationary
case? In reality, we need to give the vorticity in both cases, but in the
nonstationary case the vorticity is determined by the initial velocity field 
$\mathbf{v}_{\mathrm{in}}\left( \mathbf{x}\right) $, given in the whole
volume $V$. In the stationary case the velocity is given only on the
boundary $\partial V$, and the vorticity cannot be determined on the basis
of this information. In this case we can determine only one component of the
vorticity vector $\mathbf{\omega }=\mathbf{\nabla }\times \mathbf{v}$. As a
result the vorticity must be given in addition.

Now we consider the equations (\ref{s5.16}), which contain indefinite
functions $\Omega ^{\alpha \beta }\left( \xi _{2},\xi _{3}\right) $. As we
have seen in section 4, the functions $g^{\alpha }\left( \mathbf{\xi }%
\right) $, $\alpha =1,2,3$ contain the complete information, which is
necessary for determination of the unique fluid flow, whereas initial and
boundary conditions for variables $\mathbf{\xi }$ do not influence the fluid
flow and may be taken arbitrarily. In the case of the irrotational flow all
functions $\Omega ^{\alpha \beta }\left( \xi _{2},\xi _{3}\right) =0$, and
one obtains from (\ref{s5.16}) 
\begin{equation}
\mathbf{\nabla }\times \mathbf{v}=0,\qquad \mathbf{v}=\left( \mathbf{\nabla }%
\xi _{2}\times \mathbf{\nabla }\xi _{3}\right)   \label{s5.21}
\end{equation}
Replacing the second equation (\ref{s5.21}) by the equivalent equation $%
\mathbf{\nabla v}=0$, we reduce the problem to equations (\ref{s5.19}), (\ref
{s5.20}).

In the case of the rotational flow the vorticity $\Omega ^{\alpha \beta }$
is to be given on any stream line, labeled by $\xi _{2}$, $\xi _{3}$. If the
fluid flow contains closed stream lines, which do not cross the boundary $%
\partial V$, one cannot determine the vorticity $\Omega ^{\alpha \beta }$ on
these stream lines, giving boundary conditions. In this case one needs an
additional information other than the boundary condition. We can obtain the
stationary flow as a result of establishing process. In this case we are to
give initial velocity $\mathbf{v}_{\mathrm{in}}\left( \mathbf{x}\right) $ in
the whole volume $V$, and the vorticity on the closed stream lines is
determined by the initial velocity $\mathbf{v}_{\mathrm{in}}\left( \mathbf{x}%
\right) $. Using different establishing processes, we obtain different
stationary flows at the same stationary boundary conditions. Experimenters
dealing with rotating fluid flows know this fact very well.

Equations (\ref{s5.16}) are equations for the generalized stream function
(GSF) $\left\{ \xi _{2},\xi _{3}\right\} $. These equations are nonlinear.
In terms of the variables $\mathbf{v},\xi _{2},\xi _{3}$ equations (\ref
{s5.16}) contain terms linear with respect $\mathbf{v}$. Equations (\ref
{s5.16}) became to be simple in the special case, when $\Omega ^{23}=$const
and $\Omega ^{13}=\Omega ^{12}=0$. In this case we obtain linear equations 
\begin{equation}
\mathbf{\nabla }\times \mathbf{v}=\Omega ^{23}\mathbf{v,\qquad \nabla v}%
=0,\qquad \Omega ^{23}=\text{const}  \label{s5.23}
\end{equation}
which can be solved without consideration of variables $\xi _{2},\xi _{3}$.

\section{Stationary flow of incompressible fluid described in Lagrangian
coordinates}

Dynamic equations (\ref{s5.16}) contain indefinte functions $\Omega ^{\alpha
\beta }\left( \mathbf{\xi }\right) $ as functions of dependent variables.
Sometimes it is useful to have dynamic equations, where indefinite functions
be those of independent variables. Let us transform dynamic equations (\ref
{s5.3}), (\ref{s5.7}) and (\ref{s5.11}) to independent variables $t,\mathbf{%
\xi }$.

We introduce designations (\ref{b3.18}) 
\begin{equation}
x^{\alpha }=x^{\alpha }\left( t,\xi _{1},\xi _{2},\xi _{3}\right) ,\qquad
x^{\alpha ,\beta }\equiv \frac{\partial x^{\alpha }}{\partial \xi _{\beta }}%
,\qquad \alpha ,\beta =1,2,3  \label{h7.1}
\end{equation}

\begin{equation}
X\equiv \frac{\partial \left( x^{1},x^{2},x^{3}\right) }{\partial \left( \xi
_{1},\xi _{2},\xi _{3}\right) }=\det \left| \left| x^{\alpha ,\beta }\right|
\right| ,\qquad \alpha ,\beta =1,2,3  \label{h7.2}
\end{equation}
We consider the Jacobian $X$ as the 3-linear function of arguments $%
x^{\alpha ,\beta }$. We obtain 
\begin{equation}
\xi _{\alpha ,\beta }=X^{-1}\frac{\partial X}{\partial x^{\beta ,\alpha }}%
,\qquad \alpha ,\beta =1,2,3  \label{h7.2a}
\end{equation}
\begin{equation}
v^{\alpha }=\frac{\partial \left( x^{\alpha },\xi _{2},\xi _{3}\right) }{%
\partial \left( x^{1},x^{2},x^{3}\right) }=\frac{\partial \left( x^{\alpha
},\xi _{2},\xi _{3}\right) }{\partial \left( \xi _{1},\xi _{2},\xi
_{3}\right) }\frac{\partial \left( \xi _{1},\xi _{2},\xi _{3}\right) }{%
\partial \left( x^{1},x^{2},x^{3}\right) }=x^{\alpha ,1},\qquad \alpha =1,2,3
\label{h7.3}
\end{equation}
We write dynamic equations (\ref{s5.3}), (\ref{s5.7}) and (\ref{s5.11}) in
the form 
\begin{equation}
X\equiv \frac{\partial \left( x^{1},x^{2},x^{3}\right) }{\partial \left( \xi
_{1},\xi _{2},\xi _{3}\right) }=1  \label{h7.4}
\end{equation}
\begin{equation}
v^{\alpha }\xi _{1,\alpha }=1,\qquad v^{\alpha }\xi _{2,\alpha }=0,\qquad
v^{\alpha }\xi _{3,\alpha }=0,  \label{h7.5}
\end{equation}
\begin{equation}
\varepsilon _{\mu \alpha \beta }\partial _{\alpha }v^{\beta }=\frac{1}{2}%
\varepsilon _{\mu \alpha \beta }\Omega ^{\rho \sigma }\left( \xi _{2},\xi
_{3}\right) \xi _{\rho ,\alpha }\xi _{\sigma ,\beta }  \label{h7.6}
\end{equation}
where $\varepsilon _{\mu \alpha \beta }$ is the Levi-Chivita pseudotensor.

Substituting $\xi _{\alpha ,\beta }$ and $v^{\alpha }$, taken from (\ref
{h7.2a}), (\ref{h7.3}) in (\ref{h7.5}), we obtain equations 
\begin{equation}
x^{\alpha ,1}X^{-1}\frac{\partial X}{\partial x^{\alpha ,1}}=1,\qquad
x^{\alpha ,1}X^{-1}\frac{\partial X}{\partial x^{\alpha ,2}}=0,\qquad
x^{\alpha ,1}X^{-1}\frac{\partial X}{\partial x^{\alpha ,3}}=0,  \label{h7.7}
\end{equation}
which are satisfied identically. We substitute $\xi _{\alpha ,\beta }$ and $%
v^{\alpha }$, taken from (\ref{h7.2a}), (\ref{h7.3}) in (\ref{h7.6}). Taking
into account (\ref{h7.4}), we obtain 
\begin{equation}
x^{\beta ,1\gamma }\frac{\partial X}{\partial x^{\alpha ,\gamma }}-x^{\alpha
,1\gamma }\frac{\partial X}{\partial x^{\beta ,\gamma }}=\Omega ^{\rho
\sigma }\left( \xi _{2},\xi _{3}\right) \frac{\partial X}{\partial x^{\alpha
,\rho }}\frac{\partial X}{\partial x^{\beta ,\sigma }}  \label{h7.8}
\end{equation}
Convolving (\ref{h7.8}) with $x^{\alpha ,\mu }$ and $x^{\beta ,\nu }$and
taking into account (\ref{h7.4}), we obtain 
\begin{equation}
x^{\alpha ,1\mu }x^{\alpha ,\nu }-x^{\alpha ,1\nu }x^{\alpha ,\mu }=\Omega
^{\mu \nu }\left( \xi _{2},\xi _{3}\right) ,\qquad \mu ,\nu =1,2,3
\label{h7.9}
\end{equation}
where summation is made over $\alpha =1,2,3$.

In the three-dimensional space of coordinates $\xi _{1},\xi _{2},\xi _{3}$
the equations (\ref{h7.9}) can be written in the vector form 
\begin{equation}
\sum\limits_{\alpha =1}^{\alpha =3}\mathbf{\nabla }_{\xi }v^{\alpha }\times 
\mathbf{\nabla }_{\xi }x^{\alpha }=\mathbf{\omega }\left( \xi _{2},\xi
_{3}\right) ,\qquad \mathbf{\nabla }_{\xi }=\left\{ \frac{\partial }{%
\partial \xi _{1}},\frac{\partial }{\partial \xi _{2}},\frac{\partial }{%
\partial \xi _{3}}\right\} ,\qquad v^{\alpha }=x^{\alpha ,1},\qquad \alpha
=1,2,3  \label{h7.10}
\end{equation}
\begin{equation}
\mathbf{\omega =}\left\{ \omega _{1},\omega _{2},\omega _{3}\right\} ,\qquad
\omega _{\mu }\left( \xi _{2},\xi _{3}\right) =\frac{1}{2}\varepsilon _{\mu
\alpha \beta }\Omega ^{\alpha \beta }\left( \xi _{2},\xi _{3}\right) ,\qquad
\mu =1,2,3  \label{h7.11}
\end{equation}
where $\varepsilon _{\mu \alpha \beta }$ is the Levi-Chivita pseudotensor.

Finally, equation (\ref{h7.4}) can be resolved with respect to $x^{1}$ in
the same way, as the equation (\ref{s5.7}) has been resolved with respect to 
$\xi _{1}$. We obtain 
\begin{equation}
x^{1}=\int \frac{ds}{\Delta _{1}\left( s,x^{2},x^{3}\right) },\qquad \Delta
_{1}=\frac{\partial \left( s,x^{2},x^{3}\right) }{\partial \left( \xi
_{1},\xi _{2},\xi _{3}\right) }  \label{h7.12}
\end{equation}
where $s$ is some function of arguments $\xi _{1},\xi _{2},\xi _{3}$ and the
Jacobian $\Delta _{1}$ is considered as a function of arguments $%
s,x^{2},x^{3}$. In particular, if $s=\xi _{1}$, 
\begin{equation}
\Delta _{1}=\frac{\partial \left( x^{2},x^{3}\right) }{\partial \left( \xi
_{2},\xi _{3}\right) }=\frac{1}{v^{1}\left( \xi _{1},x^{2},x^{3}\right) }%
,\qquad x^{1}=\int v^{1}\left( \xi _{1},x^{2},x^{3}\right) d\xi _{1}
\label{h7.14}
\end{equation}

Using relation (\ref{h7.12}) or (\ref{h7.14}), we can eliminate variables $%
x^{1}$ and $v^{1}=x^{1,1}$ from equations (\ref{h7.10}). We obtain the
system of dynamic equations for variables $x^{2},x^{3}$ in the
two-dimensional space of coordinates $\xi _{2},\xi _{3}$ with variable $\xi
_{1}$, considered as an evolutional variable (time).

Equations (\ref{h7.10}), (\ref{h7.14}) are rather complicated, especially
because of the equation (\ref{h7.14}), which contains the operation of
transition from independent variables $\left\{ \xi _{1},\xi _{2},\xi
_{3}\right\} $ to independent variables $\left\{ \xi
_{1},x^{2},x^{3}\right\} $. Properties of this operation are investigated
slightly. Apparently, this operation is an attribute of incompressible
fluid, because this operation is present in dynamic equations (\ref{s5.16}),
written in the Eulerian coordinates. Indefinite functions $\Omega ^{\alpha
\beta }$ describing vorticity are functions of independent variables in the
equations (\ref{h7.10}), whereas they are functions of dependent variables
in equations (\ref{s5.16}). From this viewpoint the equations (\ref{h7.10})
are more convenient for investigation of vortical flows.

\section{Concluding remarks}

We have seen that the statement of the flow problem is more complicated for
stationary flows, than for nonstationary ones, although dynamic equations
for nonstationary flows are more complicated. Stationary boundary conditions
do not destine a unique stationary flow, even if they determine vorticity on
any stream line crossing the boundary of the considered flow region. There
is a hope that the description of stationary flows of the ideal fluid may
appear to be more effective in Lagrangian coordinates.

%\newpage

\end{document}